\documentclass[amsmath,amssymb,twocolumn]{revtex4}
\usepackage{subfig}
\usepackage{dcolumn}
\usepackage{bm}
\usepackage{amsmath,mathrsfs}
\usepackage{graphicx,epsfig}
\usepackage{hyperref}
\usepackage{epsf}
\usepackage{color}

\begin{document}


\title{{\bf Wormhole Geometries in Extended Symmetric Teleparallel Gravity with $f(Q, T) = \alpha Q + \beta T + \gamma T^2$}}

\author{S. Rastgoo}\email{rastgoo@sirjantech.ac.ir}
\author{ F. Parsaei }\email{fparsaei@gmail.com}

\affiliation{ Physics Department , Sirjan University of Technology, Sirjan 78137, Iran.}



\begin{abstract}
\par   In this paper, we explore static and spherically‑symmetric wormhole configurations within the recently proposed class of extended symmetric teleparallel gravities characterised by the functional form $f(Q,T)=\alpha Q+\beta T+\gamma T^{2}$, where $Q$ is the non‑metricity scalar, $T$ is the trace of the  energy‑momentum tensor, and ($\alpha,\beta,\gamma$) are constant coupling parameters. By varying the action with respect to the metric, we derive the modified field equations and cast them in an effective Einstein‑like form containing additional geometric contributions that depend on $Q$ and $T$.
We conclude that the $f(Q,T)=\alpha Q+\beta T+\gamma T^{2}$ framework offers a viable arena for constructing physically realistic wormholes without invoking severe energy condition violations, opening new pathways for connecting modified gravity phenomenology with astrophysical signatures of non‑trivial spacetime topologies.\\
\end{abstract}

\maketitle

\section{Introduction}

A wormhole is a theoretical tunnel-like solution to the equations governing gravitational fields, linking two distinct regions of spacetime (or two separate Universes) \cite{WH}. In general relativity (GR), static, spherically symmetric wormholes (known as the Morris–Thorne class) necessitate the presence of exotic matter that contravenes the null energy condition (NEC) at the throat \cite{Visser}. Because exotic matter is generally regarded as improbable within GR, many investigators have sought ways to lessen or even eliminate its required presence. Their approaches aim to satisfy the geometric conditions of a wormhole while reducing dependence on exotic matter, confining any such matter to limited regions of spacetime. Notable models that embody this idea include thin-shell wormholes \cite{cut, cut2, cut3}, wormholes with a variable equation-of-state (EoS)\cite{Remo, variable}, and wormholes described by a polynomial EoS \cite{foad}.
Following the detection of the Universe’s accelerated expansion, researchers have turned their attention to studying phantom wormholes in the scientific literature \cite{phantom, phantom2, phantom1}.

The emergence of extended modified gravities, led to significant investigation into the possibility that modified theories of gravity might accommodate wormhole geometries without the need for matter that blatantly breaches the standard energy conditions (ECs). The additional geometric degrees of freedom can effectively provide the required 'exoticity,' enabling ordinary or minimally exotic matter to traverse the throat. Wormholes have been investigated within a broad spectrum of modified‑gravity frameworks, such as the Braneworld models \cite{b1, b2, b3},  quadratic gravity \cite{quad, quad1},    $f(Q)$ gravity \cite{fq, fq2, fq4, fq44, fq5}, $f(R)$ gravity \cite{Nojiri, fR11, fR22, fR55}, Ricci inverse gravity \cite{inverse}, Einstein-Cartan gravity \cite{Cartan, Cartan1, Cartan2, Cartan3},  Rastall gravity \cite{Rast, Rast1}, Rastall–Rainbow gravity \cite{RaR, RaR1}, $f(T,\mathcal{T})$ gravity \cite{Must1, Err, Riz, Must2, foad3}, $f(R, L_m)$ gravity \cite{L1, L2, L7, L10, L12, SR2} and $f(R,T)$ gravity \cite{ Moa, Zub, fr2, Sha, Ban, Sarkar, SR1, foad4}. In each of these settings, the Einstein field equations are modified, and the resulting alterations can, in certain cases, eliminate the need for exotic matter in wormhole constructions.

$f(Q)$ gravity is a class of modified theories of gravitation in which the fundamental geometric quantity driving the dynamics is the non-metricity scalar $Q$, rather than curvature (as in GR) or torsion (as in teleparallel gravity) \cite{f(Q), Hei}. In the so‑called symmetric‑teleparallel formulation of gravity the Levi‑Civita connection is replaced by a flat, torsion‑free connection whose only non‑vanishing piece is the non‑metricity tensor, and the standard Einstein-Hilbert action ($R$) is recast as a linear function of $Q$. By promoting this linear dependence to an arbitrary function $f(Q)$, one obtains field equations that can mimic dark energy, generate late‑time cosmic acceleration, and address cosmological tensions while preserving second‑order differential equations and avoiding ghosts. Consequently, $f(Q)$ gravity has become a popular arena for exploring viable extensions of GR that remain close to the metric‑compatible limit yet offer rich phenomenology for both cosmology and astrophysical tests.

A natural extension of the pure $f(Q)$ theory is to allow an explicit coupling between geometry and matter through the trace ($T$) of the energy‑momentum tensor (EMT) \cite{fqt}. The most general low‑order functional form that captures this coupling while remaining analytically tractable  is  $f(Q,T)= \alpha Q + \beta T$.
  Recent studies have investigated wormhole configurations within the linear formulation of $f(Q,T)$ gravity,  yielding a substantial body of literature on the subject \cite{5, 6, 7, 9, 11, 12a, 1, 12, 2, 3, 4, 8, 13, 10, 14, fq-sara, Sodab, 15, 16}. Recently, we demonstrated that non‑exotic  asymptotically flat wormhole solutions can be realized within the framework of $f(Q,T)=\alpha Q+\beta T$ gravity \cite{fq-sara, Sodab}. In particular, we examined how the free parameters $\alpha$ and $\beta$ affect the satisfaction of the various ECs in wormhole models. Now, we will investigate the possible wormhole solutions in the
context of $f(Q,T)=\alpha Q+ \beta T + \gamma T ^2$ gravity. This model extends symmetric teleparallel gravity by introducing linear and quadratic couplings between the spacetime non‑metricity and the trace of the matter EMT.

The remainder of this manuscript is structured as follows: Section \ref{sec2} introduces the concept of wormholes and provides an overview of $f(Q,T)$ gravity in relation to the classical ECs. Section \ref{sec3} explores solutions that satisfy these ECs, while Section \ref{sec4} summarizes our primary findings. Throughout this work, we employ gravitational units such that $c = 8\pi G = 1$."

\section{Basic formulation of wormhole and $f(Q,T)$ gravity} \label{sec2}
The Morris–Thorne wormhole metric, which is static and spherically symmetric, can be written as follows
\begin{equation}\label{1}
ds^2=-U(r)dt^2+\frac{dr^2}{1-\frac{b(r)}{r}}+r^2(d\theta^2+\sin^2\theta
d\phi^2),
\end{equation}
where $U(r)=\exp2\Phi(r)$. In this context, $\Phi(r)$ is defined as the redshift function, while $b(r)$ is referred to as the shape function. The redshift function governs how the gravitational redshift manifests, while the shape function defines the wormhole’s spatial geometry.
The wormhole’s throat acts as a bridge that connects either two distinct Universes or two far‑apart regions within a single Universe. The condition governing the throat is
\begin{equation}\label{2}
b(r_0)=r_0,
\end{equation}
where $r_{0}$ is the throat of wormhole. The conditions listed below must be met,
\begin{equation}\label{3}
b'(r_0)<1
\end{equation}
and
\begin{equation}\label{4}
b(r)<r,\ \ {\rm for} \ \ r>r_0.
\end{equation}
Equation (\ref{3}) is commonly called the flaring‑out condition; it signals that the NEC is violated in the framework of GR. In addition, any viable wormhole solution must be asymptotically flat so that it matches the observed large-scale behavior of the Universe,
\begin{equation}\label{5}
\lim_{r\rightarrow \infty}\frac{b(r)}{r}=0,\qquad   \lim_{r\rightarrow \infty}U(r)=1.
\end{equation}
To eliminate tidal forces, the redshift function must be constant. For simplicity, we adopt the commonly used case of a vanishing redshift function. In the present study, we consider an anisotropic fluid whose EMT  is  $T_{\nu}^{\mu}=diag[-\rho, p, p_{t}, p_{t}]$ where $\rho$ denotes the energy density, $p$ the radial pressure, and $p_t$ the transverse (tangential) pressure. Continuing, we present a concise overview of $f(Q,T)$ gravity.

The action of $f(Q, T)$ gravity is given as
\begin{equation}\label{6a}
S=\int\frac{1}{2}\,f(Q,T)\sqrt{-g}\,d^4x+\int L_m\,\sqrt{-g}\,d^4x\, ,
\end{equation}
where $f(Q,T)$ denotes an arbitrary function of the non‑metricity scalar ($Q$) and the trace ($T$) of the EMT, $\sqrt{-g}$ is the square root of the determinant of the metric tensor, and $L_{m}$ stands for the matter Lagrangian density \cite{6}. The non-metricity tensor is given by\cite{6}
\begin{equation}\label{6ab}
Q_{\lambda\mu\nu}=\bigtriangledown_{\lambda} g_{\mu\nu},
\end{equation}
Moreover, the superpotential (also called the non‑metricity conjugate) is defined as
\begin{eqnarray}\label{6ab1}
P^\alpha\;_{\mu\nu}=\frac{1}{4}\Bigg[&-&Q^\alpha\;_{\mu\nu}+2Q_{(\mu}\;^\alpha\;_{\nu)}\nonumber \\
&+&Q^\alpha g_{\mu\nu}-\tilde{Q}^\alpha g_{\mu\nu}-\delta^\alpha_{(\mu}Q_{\nu)}\Bigg],
\end{eqnarray}
where
\begin{equation}\label{a1}
Q_{\alpha}=Q_{\alpha}\;^{\mu}\;_{\mu},\; \tilde{Q}_\alpha=Q^\mu\;_{\alpha\mu},
\end{equation}
are two traces of the non-metricity tensor. The scalar that characterizes non‑metricity, denoted by\cite{6}
\begin{eqnarray}\label{a2}
Q &=& -Q_{\alpha\mu\nu}\,P^{\alpha\mu\nu} \nonumber \\
&=& -g^{\mu\nu}\left(L^\beta_{\,\,\,\alpha\mu}\,L^\alpha_{\,\,\,\nu\beta}-L^\beta_{\,\,\,\alpha\beta}\,L^\alpha_{\,\,\,\mu\nu}\right),
\end{eqnarray}
with
\begin{equation}\label{a3}
L^\beta_{\,\,\,\mu\nu}=\frac{1}{2}Q^\beta_{\,\,\,\mu\nu}-Q_{(\mu\,\,\,\,\,\,\nu)}^{\,\,\,\,\,\,\beta}.
\end{equation}
For the metric given in (\ref{1}), the non-metricity scalar is expressed as follows:
\begin{equation}\label{aa3}
Q=-\frac{b}{r^2}\left[\frac{rb^{'}-b}{r(r-b)}\right].
\end{equation}
Now, by applying the variational principle to the action (\ref{6a}), the field equations of the $f(Q,T)$ gravity theory  are derived as follows:
 \begin{multline}\label{6b}
\frac{-2}{\sqrt{-g}}\bigtriangledown_\alpha\left(\sqrt{-g}\,f_Q\,P^\alpha\;_{\mu\nu}\right)-\frac{1}{2}g_{\mu\nu}f \\
+f_T \left(T_{\mu\nu} +\Theta_{\mu\nu}\right)\\
-f_Q\left(P_{\mu\alpha\beta}\,Q_\nu\;^{\alpha\beta}-2\,Q^
{\alpha\beta}\,\,_{\mu}\,P_{\alpha\beta\nu}\right)= T_{\mu\nu},
\end{multline}
where $f_{Q}=\frac{\partial f}{\partial Q}$ and $f_{T}=\frac{\partial f}{\partial T}$.
One can write the EMT  as
 \begin{equation}\label{a4}
T_{\mu\nu}=-\frac{2}{\sqrt{-g}}\frac{\delta\left(\sqrt{-g}\,L_m\right)}{\delta g^{\mu\nu}},
\end{equation}
and
\begin{equation}\label{a5}
\Theta_{\mu\nu}=g^{\alpha\beta}\frac{\delta T_{\alpha\beta}}{\delta g^{\mu\nu}}.
\end{equation}
We assume that the $L_m$ depends solely on the metric itself and not on its derivatives.
The explicit form of the field equations hinges on the particular choices made for the function $f(Q,T)$ and for the matter Lagrangian $L_{m}$. In Lagrangian‑based gravity theories one usually adopts one of three conventional expressions for the matter Lagrangian: $[ L_{m}= \rho,\qquad L_{m}= -T ,\qquad L_{m}= -P]$, where the trace of the EMT is $T = -\rho + p + 2p_{t}$ and the average pressure is $ P = \frac{p + 2p_{t}}{3}$.  In the next section, we will work with the choice $L_{m}= -P$ and assume a non-linear form for $f(Q,T)$ to construct wormhole solutions.

ECs are sets of inequalities imposed on the EMT in GR that encode physically reasonable expectations about matter and energy. The NEC requires $T_{ab}k^{a}k^{b}\ge0$ for every null vector $k^{a}$, guaranteeing that light rays are not defocused by exotic matter. The weak energy condition (WEC) demands $T_{ab}v^{a}v^{b}\ge0$ for all timelike observers $v^{a}$, ensuring that every observer measures a non‑negative local energy density. The dominant energy condition (DEC) strengthens the WEC by also insisting that the energy flux never exceeds the speed of light, i.e. $T_{ab}v^{a}$ is a future‑directed causal vector. The strong energy condition (SEC) imposes $T_{ab}-\tfrac{1}{2}T g_{ab})v^{a}v^{b}\ge0$ for timelike $v^{a}$, which, through the Raychaudhuri equation, leads to the attractive nature of gravity used in singularity theorems. Wormhole solutions in the framework of GR have been shown to violate ECs. The simple forms of ECs are
\begin{eqnarray}\label{21}
\textbf{NEC}&:& \rho+p\geq 0,\quad \rho+p_t\geq 0 \\
\label{21a}
\textbf{WEC}&:& \rho\geq 0, \rho+p\geq 0,\quad \rho+p_t\geq 0, \\
\textbf{DEC}&:& \rho\geq 0, \rho-|p|\geq 0,\quad \rho-|p_t|\geq 0, \\
\textbf{SEC}&:& \rho+p\geq 0,\, \rho+p_t\geq 0,\rho+p+2p_t \geq 0. \label{21b}
\end{eqnarray}
These conditions are central to many classic results—such as the Hawking–Penrose singularity theorems, black‑hole area theorems, and the positivity of mass—yet they are known to be violated by quantum fields (e.g., Casimir effect) and certain cosmological models (e.g., dark energy with $\omega<-1$), prompting the development of averaged or quantum‑energy conditions as more flexible alternatives. According to  \cite{fq}, we will investigate the ECs by defining the functions,
\begin{eqnarray}\label{22}
 H(r)&=& \rho+p ,\, H_1(r)= \rho+p_t,\, H_2(r)= \rho-|p|, \nonumber \\
 H_3(r)&=&\rho-|p_t|,\, H_4(r)= \rho+p+2p_t .
\end{eqnarray}

A standard method for deriving exact wormhole solutions involves using field equations to determine the shape function from a given EoS. Alternatively, some researchers define shape functions with free parameters, adjusting them to satisfy specific physical and mathematical criteria. In the following section, we employ the former approach to identify asymptotically flat wormhole solutions, setting $r_0 = 1$ for simplicity.

\section{Non-exotic wormhole solutions }\label{sec3}
The linear form for $f(Q,T)$ is
\begin{equation}\label{99a}
f(Q, T)=\alpha  Q+\beta T,
\end{equation}
where $\alpha$ and $\beta$ are free parameters. This is the primary and most straightforward model explored in the literature \cite{5, 6, 7, 9, 11, 12a, 1, 12, 2, 3, 4, 8, 13, 10, 14, fq-sara, Sodab, 15, 16}. The next candidate is
\begin{equation}\label{F1}
f(Q, T)=\alpha  Q+\beta T+ \gamma T^2,
\end{equation}
where a term proportional to $T^2$ is added to the linear model. This additional term significantly alters the field equations, diverging substantially from the linear model (\ref{99a}). In this research, we adopt the $f(Q,T)$ in the form of (\ref{F1}) and seek possible wormhole solutions.
By using metric (\ref{1}) and Eqs.(\ref{6ab}-\ref{a5}), one can reach  the following field equation
\begin{eqnarray}\label{19aa}
\rho = \frac{(r-b)}{2 r^3}  \Bigg[f_Q \left(\frac{(2 r-b) (r b'-b)}{(r-b)^2}+\frac{2b}{r-b}\right)
 \nonumber \\
+\frac{2 b r f_{\text{QQ}} Q'}{r-b}+\frac{f r^3}{r-b}-\frac{2r^3 f_T (-L_m+\rho )}{(r-b)} \Bigg],
\end{eqnarray}
\begin{eqnarray}\label{11cc}
 p = -\frac{(r-b)}{2 r^3} \Bigg[f_Q \frac{b }{r-b}\left(\frac{r b'-b}{r-b}+2\right) \nonumber \\
 +\frac{2 b r f_{\text{QQ}} Q'}{r-b}+\frac{f r^3}{r-b}-\frac{2r^3 f_T (-L_m-p)}{(r-b)}\Bigg],
 \end{eqnarray}
 \begin{eqnarray}\label{11d}
 p_t = -\frac{(r-b)}{4 r^2} \Bigg[2f_Q \left(\frac{r b'-b}{(r-b)^2}\right)+\frac{2 f r^2}{r-b}\nonumber \\
 -\frac{4r^2 f_T (-L_m-p_t)}{(r-b)}\Bigg].
 \end{eqnarray}
Using Eqs.(\ref{F1}) and (\ref{aa3}) in Eqs.(\ref{19aa}-\ref{11d}) gives
\begin{eqnarray}\label{3e}
  \rho =\frac{1}{2 r^3} \Bigg[\alpha \left(\frac{(2 r-b) \left(r b'-b\right)}{(r-b)}+2b\right) \nonumber \\
 +f r^3+4r^3 \gamma\, T (-P-\rho )\Bigg],
\end{eqnarray}
\begin{eqnarray}\label{3f}
 p=-\frac{1}{2 r^3} \Bigg[\alpha b\left(\frac{r b'-b}{r-b}+2\right)\nonumber \\
 +f r^3+4r^3 \gamma\, T \left(-P+p\right)\Bigg],
\end{eqnarray}
\begin{eqnarray}\label{3g}
  p_t=-\frac{1}{2 r^2} \Bigg[\alpha \left(\frac{\left(r b'-b\right) }{r-b }\right)+f r^2\nonumber \\
  -4r^2 \gamma\,T \left(P-p_t\right)\Bigg].
\end{eqnarray}
Several strategies exist for solving the field equations. Given that the three equations (Eqs. \ref{3e}–\ref{3g}) involve four unknown functions—$b(r)$, $\rho(r)$, $p(r)$, and $p_t(r)$—the system is currently underdetermined. To close the system, one may prescribe an EoS, $p(\rho) = F(\rho)$, or assume a specific energy density distribution, as seen in \cite{2, 3}. Alternatively, as demonstrated in \cite{4, 6}, one may manually specify the redshift function, $\phi(r)$, and the shape function, $b(r)$. In this section, we adopt an EoS and derive solutions that satisfy the ECs.

We assume
\begin{equation}\label{9a}
p (r)=\omega_r\rho
\end{equation}
and
\begin{equation}\label{9b}
p_t (r)=\omega_t\rho,
\end{equation}
where $\omega_r$ and $\omega_t$ are the constant EoS parameters.
In the context of modified gravity and cosmology, the linear EoS serves as a fundamental tool for modeling the dynamics of the Universe. In theories like $f(Q,T)$ gravity, this linear relationship simplifies the field equations by providing a direct link between the geometric properties of spacetime and the distribution of matter. By assuming a constant EoS parameter, researchers can investigate specific astrophysical phenomena, such as the viability of wormhole geometries or the expansion history of the Universe, by testing whether the matter-energy content satisfies or violates various ECs. This approach allows for a clean separation between the geometric modifications of the gravitational theory and the thermodynamic behavior of the cosmic fluid, making it an essential baseline for assessing whether a given gravity model can successfully explain observed cosmic acceleration or provide stable structural solutions.
Substituting the expressions for $P$ and $T$ in terms of energy density and pressure into Eqs. (\ref{3e}-\ref{3g}), and subsequently applying (\ref{9a}) and (\ref{9b}), yields
 \begin{eqnarray}
\frac{b^{\prime}(r)}{r^2} &=& \frac{a_1}{\alpha}\rho+\frac{\gamma}{\alpha}a_2\rho^2,\label{18a}\\
\frac{b(r)}{r^3} &=& \frac{a_3}{\alpha}\rho+\frac{\gamma}{\alpha}a_4\rho^2,\label{18b}\\
\frac{b^{\prime}(r)}{r^2}-\frac{b(r)}{r^3} &=& \frac{a_5}{\alpha}\rho+\frac{\gamma}{\alpha}a_6\rho^2,\label{18c}
\end{eqnarray}
where
\begin{eqnarray}
a_1 &=& 1+ \frac{\beta}{2}+\frac{5}{6}\beta\omega_r+\frac{5}{3}\beta\omega_t, \nonumber\\
a_2 &=& \frac{3}{2}-\frac{7}{6}\omega_r^2-\frac{14}{3}\omega_t^2-\frac{14}{3}\omega_r\omega_t-\frac{\omega_r}{3}-\frac{2}{3}\omega_t, \nonumber \\
a_3 &=& -\omega_r-\frac{\beta}{6}\omega_r-\frac{\beta}{2}+\frac{5\beta}{3}\omega_t, \nonumber \\
a_4 &=& -\frac{1}{2} +\frac{5}{6}\omega_r^2-\frac{14}{3}\omega_t^2-\frac{2}{3}\omega_r\omega_t-\frac{\omega_r}{3}+\frac{10}{3}\omega_t,\nonumber\\
a_5 &=& -\omega_t+\frac{2\beta}{3}\omega_t-\frac{\beta}{2}+\frac{5}{6}\omega_r, \nonumber \\
a_6 &=&-\frac{1}{2} +\frac{7}{6}\omega_r^2-\frac{2}{3}\omega_t^2-\frac{8}{3}\omega_r\omega_t+\frac{5\omega_r}{3}+\frac{4}{3}\omega_t.  \label{20a}
\end{eqnarray}
Equations (\ref{18a}-\ref{18c}) are second-order non-linear field equations that determine the solutions for $\rho$. To obtain these solutions, one must ensure consistency within the system of equations. Equations (\ref{18a}-\ref{18c}) are consistent while
\begin{eqnarray}
\frac{a_1-a_3}{\alpha} &=& \frac{a_5}{\alpha},\label{20ba}\\
\frac{\gamma}{\alpha}(a_2-a_4) &=& \frac{\gamma}{\alpha} a_6.\label{20ca}
\end{eqnarray}
It should be noted that Eqs.(\ref{20ba}) and (\ref{20ca}) must hold for $\alpha\neq0$ and $\gamma\neq0$ so
\begin{eqnarray}
a_1-a_3 &=& a_5,\label{20b}\\
a_2-a_4 &=& a_6.\label{20c}
\end{eqnarray}

 \begin{table*}[]
\begin{tabular}{|l|l|l|l|l|}
\hline
$\beta$  & $m$ & $\omega_r$  & $\omega_t$ & asymptotically flat \\ \hline
 0.8022  & $1$ & -1 & -2.65 & $\times$ \\ \hline
2.74765 & $0.9965$ & -1.011 & 2.61 &   \checkmark \\ \hline
2.773 &  $1$ & -1 & 2.615 & $\times$ \\ \hline
 -0.4765 & $1$ & -1 & -0.45 & $\times$  \\ \hline
\end{tabular}
\caption{Possible solutions for Eq. (\ref{s15}) assuming $\omega^+$.  }\label{Tab1}
\end{table*}

\begin{table*}[]
\begin{tabular}{|l|l|l|l|l|}
\hline
$\beta$  & $m$ & $\omega_r$  & $\omega_t$ & asymptotically flat \\ \hline
1.48263   & $0.3039$ & -1.9596 & 0.6961 &   \checkmark \\ \hline
-0.47520 &  $-1.7657$ & 0.6546 & -0.0647 & \checkmark\\ \hline
 -0.75309 & $-7.686$ & 0.222& 0.1729 & \checkmark  \\ \hline
  -8.34605 & $1$ & -1 & -0.508& $\times$  \\ \hline
\end{tabular}
\caption{Possible solutions for Eq. (\ref{s15}) assuming $\omega^-$.  }\label{Tab2}
\end{table*}

One can use Eqs.(\ref{18a}) and (\ref{18b}) to find
\begin{equation}\label{24d}
 \rho_{1}^{\pm}=\frac{-\frac{a_1}{\alpha}\pm\sqrt{\frac{a_1^{2}}{\alpha^2}+4\frac{\gamma}{\alpha} a_2\frac{b'}{r^2} }}{2\frac{\gamma}{\alpha}a_2}
\end{equation}
and
\begin{eqnarray}\label{25}
  \rho_{2}^{\pm}=\frac{-\frac{a_3}{\alpha}\pm\sqrt{\frac{a_3^{2}}{\alpha^2}+4\frac{\gamma}{\alpha} a_4\frac{b}{r^3} }}{2\frac{\gamma}{\alpha}a_4}.
\end{eqnarray}
For the two expressions to describe the same physical solution, the corresponding branches must be mutually consistent. For the generic case $\alpha\neq0$, $\gamma\neq0$, and non-vanishing matter density, this consistency is obtained when the two coefficient pairs are proportional, namely
 \begin{equation}\label{26}
\frac{a_1}{a_3}=\frac{a_2}{a_4}=m.
\end{equation}
Under this condition, the right-hand side of the Eq.(\ref{24d}) becomes $m$ times the right-hand side of the Eq.(\ref{25}). Consequently,
\begin{equation}\label{27a}
\frac{b'}{b}=\frac{m}{r}.
\end{equation}
Integration then gives
\begin{equation}\label{27ab}
b(r)=C r^m.
\end{equation}
Using the throat condition $b(r_0)=r_0$ and the choice $r_0=1$, we obtain $C=1$, and therefore
\begin{equation}\label{27}
b(r)=r^m.
\end{equation}
Among the various shape functions used in wormhole theory, the  power-law shape function is the most widely recognized. It remains a valid candidate for wormhole models as long as the constraint $m<1$ holds. Equation(\ref{20b}) yields
\begin{equation}\label{SS1}
1+\frac{3}{2}\beta+\omega_{r}(\beta+\frac{1}{6})+\omega_{t}(-\frac{2}{3}\beta+1)=0.
 \end{equation}
This equation implies that
\begin{equation}\label{18bb}
 \omega_r=c_1+c_2\omega_t,
 \end{equation}
where
\begin{eqnarray}
c_1=-\frac{6+9\beta}{1+6\beta}, \nonumber\\
c_2=\frac{-6+4\beta}{1+6\beta}.\label{21c}
\end{eqnarray}

Equation (\ref{20c}) yields
 \begin{equation}\label{s9}
 \frac{3}{2}-\frac{5}{6}\omega_{r}^{2}-\frac{4}{3}\omega_{r}\omega_{t}-\frac{16}{3}\omega_{t}+\frac{2}{3}\omega_{t}^{2}-\frac{5}{3}\omega_{r}=0,
 \end{equation}
Putting (\ref{18bb}) in (\ref{s9}) gives
 \begin{eqnarray}\label{s10}
\frac{3}{2}-\frac{5}{6}c_{1}^{2}-\frac{5}{3}c_{1}+(-\frac{5}{3}c_{1}c_{2}-\frac{4}{3}c_{1}-\frac{16}{3}-\frac{5}{3}c_{2})\omega_{t} \nonumber \\ +(-\frac{5}{6}c_{2}^{2}-\frac{4}{3}c_{2}+\frac{2}{3})\omega_{t}^{2}=0
 \end{eqnarray}
Solutions for (\ref{s10}) are
 \begin{equation}\label{s11}
 \omega_{t}^{\pm}=\frac{-b\pm \sqrt{b^{2}-4ac}}{2a}.
\end{equation}
where
 \begin{equation}\label{s12}
a=-\frac{5}{6}c_{2}^{2}-\frac{4}{3}c_{2}+\frac{2}{3},
\end{equation}
\begin{equation}\label{s13}
b=-\frac{4}{3}c_{1}-\frac{16}{3}-\frac{5}{3}c_{2}-\frac{5}{3}c_{2}c_{1},
\end{equation}
and
\begin{equation}\label{s14}
c=\frac{3}{2}c_{1}-\frac{5}{6}c_{1}^{2}-\frac{5}{3}c_{1}.
\end{equation}

Substituting (\ref{s11}) into (\ref{26}) yields two separate equations for the positive and negative signs. For $\omega^{\pm}_{t}$, the resulting equation is:
   \begin{eqnarray}\label{s15}
   \Bigg[ 1+\frac{\beta}{2}+\frac{5}{6}\beta c_{1}+(\frac{5}{6}\beta c_{2}
+\frac{5}{3}\beta)\omega_{t}^{\pm}\Bigg]\times\nonumber \\
\Bigg[-\frac{1}{2}+\frac{5}{6}c_{1}^{2}-\frac{1}{3}c_{1}
+(\frac{5}{3}c_{1}c_{2}-\frac{2}{3}c_{1}-\frac{1}{3}c_{2}+\frac{10}{3})\omega_{t}^{\pm}\nonumber \\
+(\frac{5}{6}c_{2}^{2}
-\frac{14}{3}-\frac{2}{3}c_{2})(\omega_{t}^{\pm})^{2}\Bigg] = \nonumber \\
\Bigg[(\frac{3}{2}-\frac{7}{6}c_{1}^{2}-\frac{1}{3}c_{1}+(\frac{7}{3}c_{1}c_{2}
-\frac{14}{3}c_{1}-\frac{1}{3}c_{2}-\frac{2}{3})\omega_{t}^{\pm} \nonumber \\
+(-\frac{7}{6}c_{2}^{2}-\frac{14}{3}-\frac{14}{3}c_{2})(\omega_{t}^{\pm})^{2}\Bigg] \nonumber \\
\times\Bigg[-c_{1}-\frac{1}{6}\beta c_{1}-\frac{\beta}{2}+(-c_{2}-\frac{1}{6}\beta c_{2}+\frac{5}{3}\beta)\omega_{t}^{\pm}\Bigg]
 \end{eqnarray}

\begin{figure}
\subfloat[ $\rho^+<0$ ]{\includegraphics[width = 2.5in]{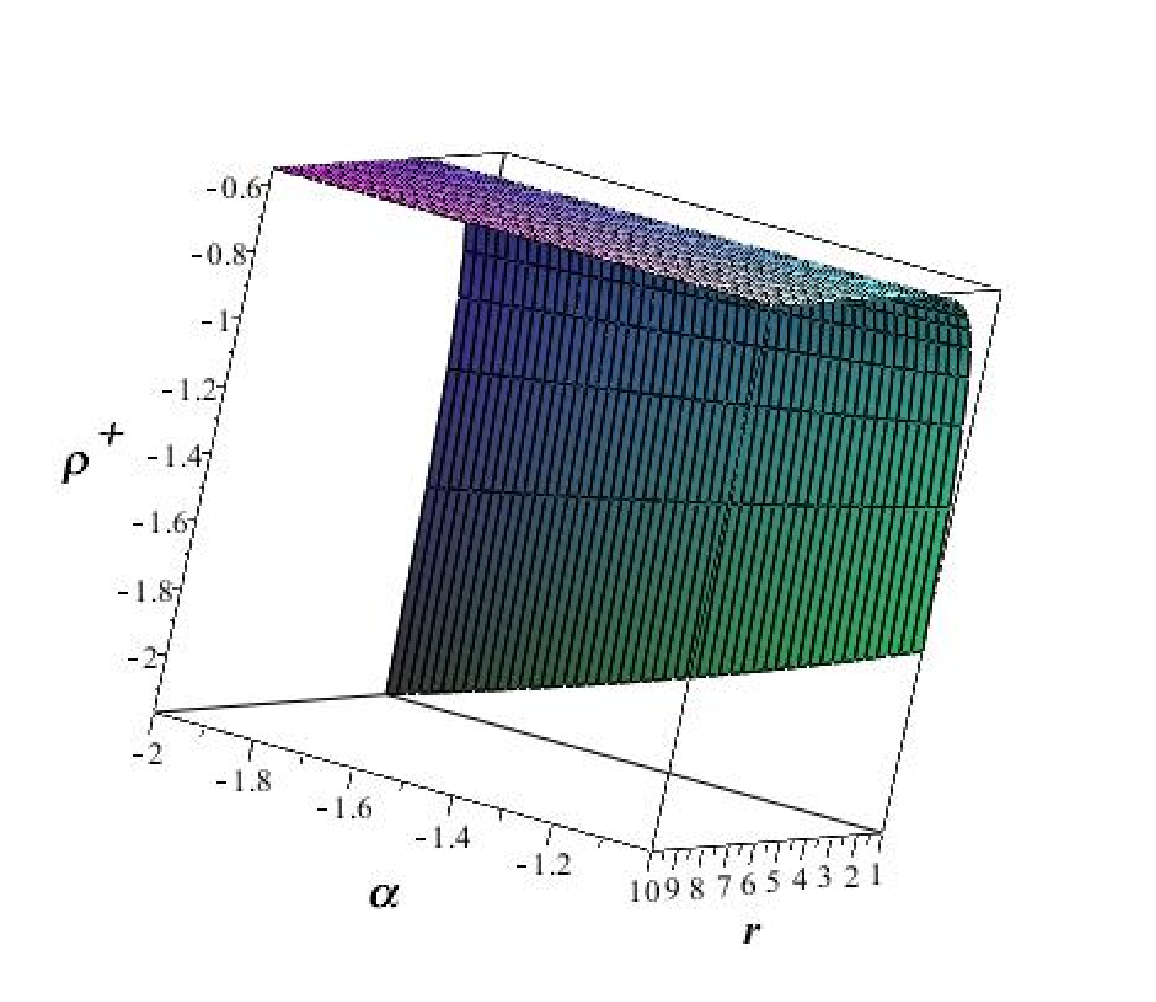}}\\
\subfloat[ $\rho^->0$]{\includegraphics[width = 2.5in]{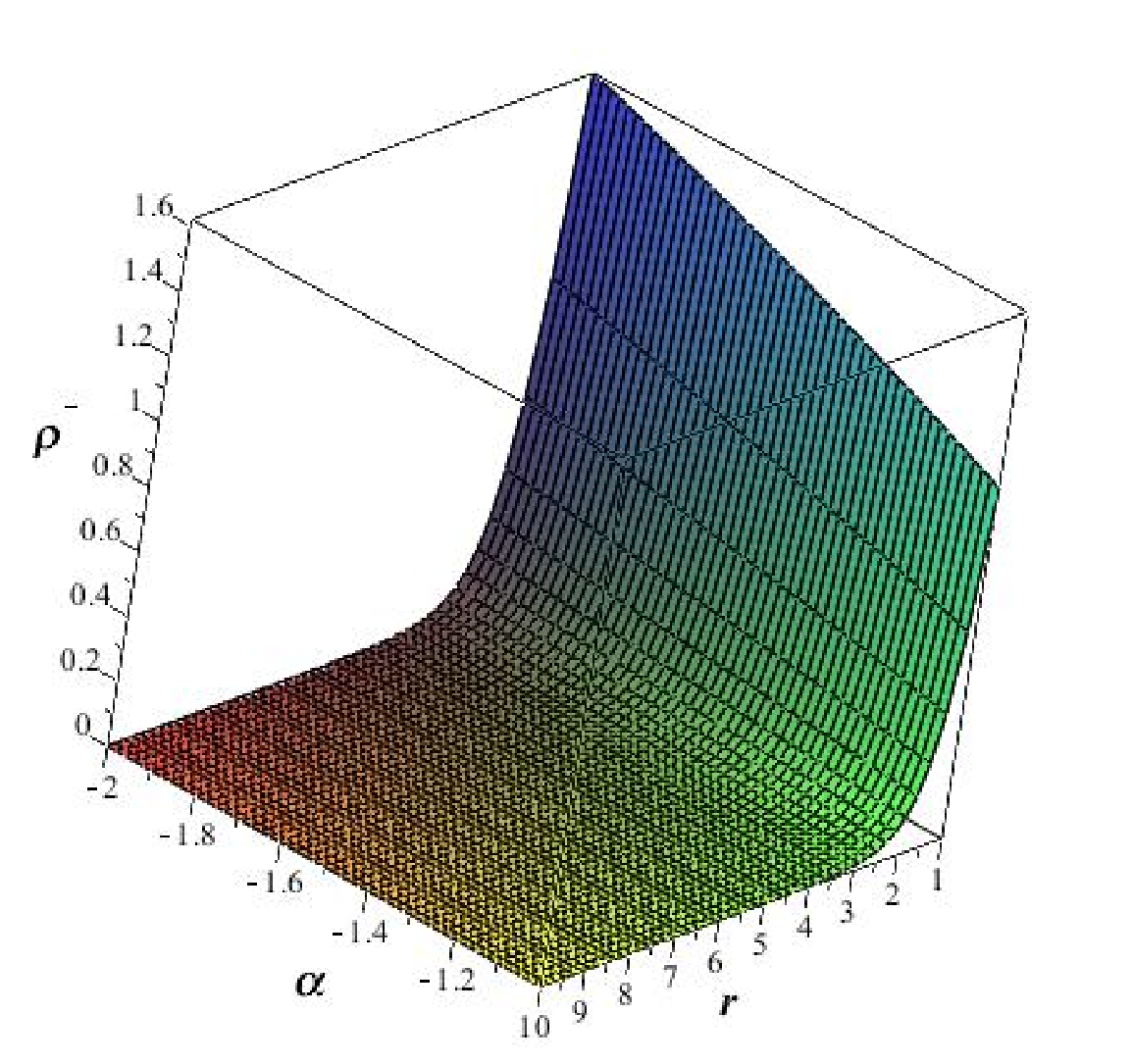}}
\caption{The relationship between $\rho^{\pm}(\alpha, r)$ and the variables $\alpha$ and $r$ for $\beta=-0.47507$ and $m=-1.7657$. Panel (a) shows the region where $\rho^+(\alpha, r) < 0$, while panel (b) illustrates the corresponding region for $\rho^-(\alpha, r)>0$.}
\label{fig1}
\end{figure}

\begin{figure}
\subfloat[ $\rho^+<0$ ]{\includegraphics[width = 2.5in]{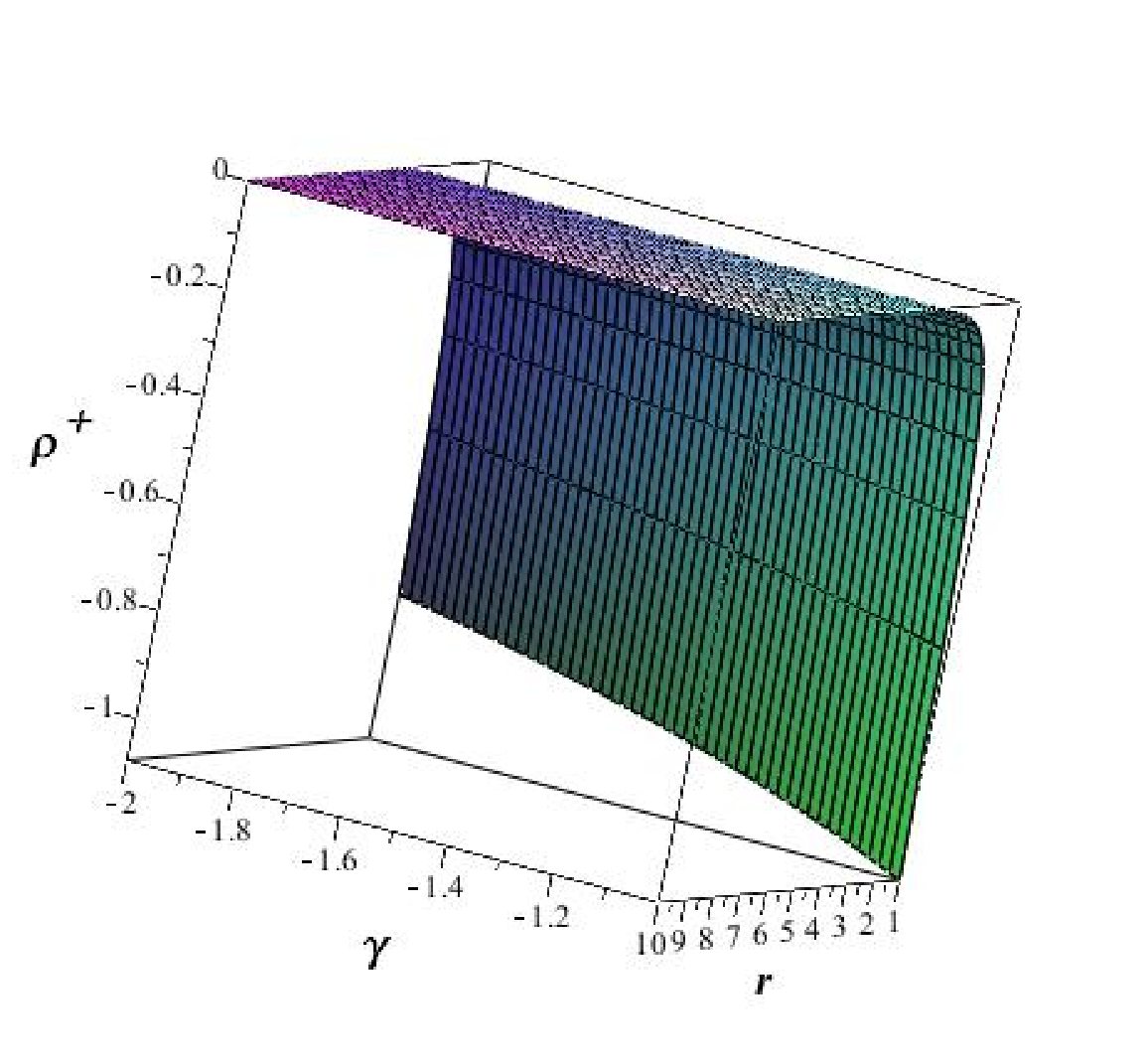}}\\
\subfloat[  $\rho^->0$]{\includegraphics[width = 2.5in]{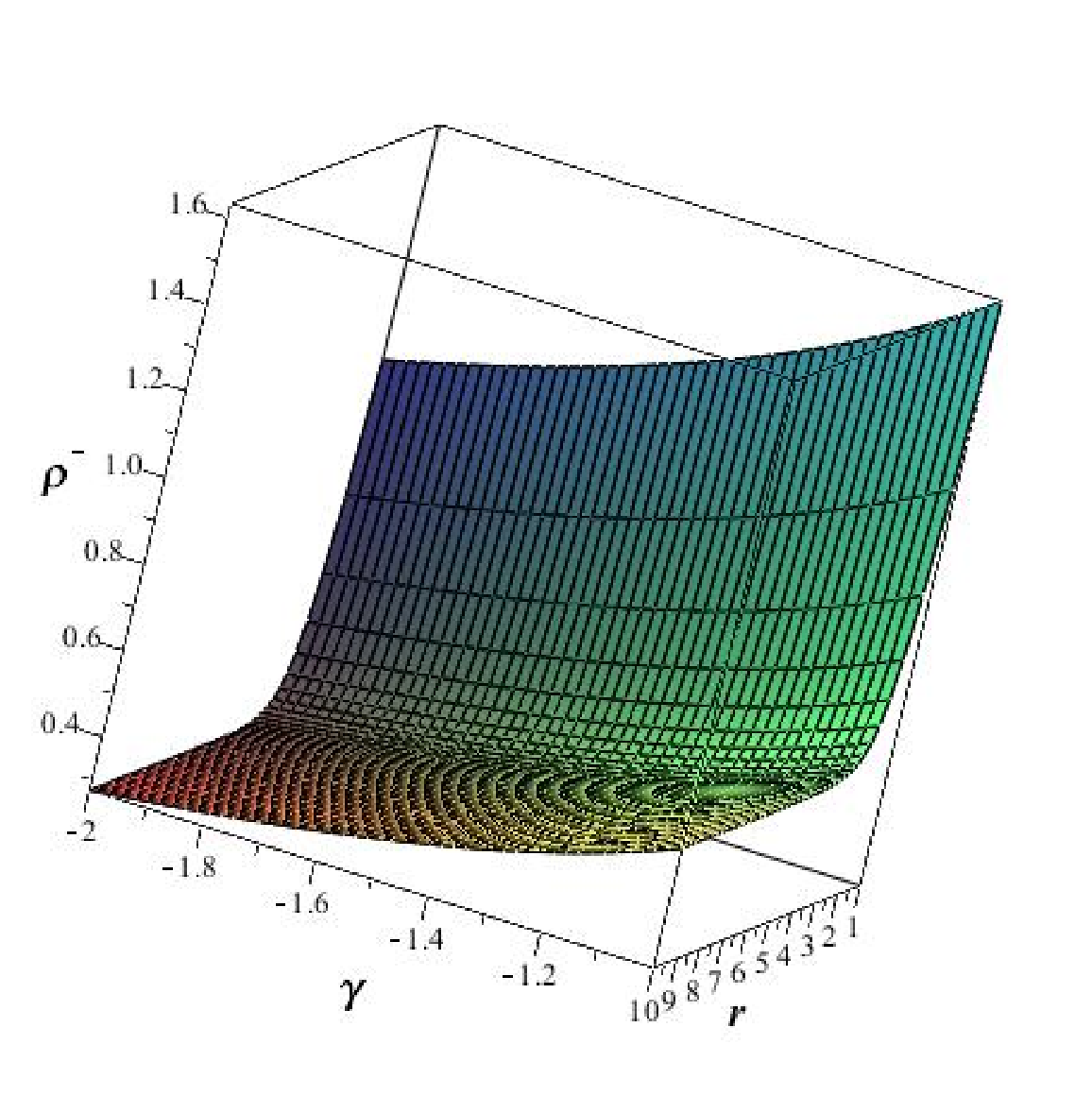}}
\caption{The relationship between $\rho^{\pm}(\gamma, r)$ and the variables $\gamma$ and $r$ for $\beta=-0.47507$ and $m=-1.7657$. Panel (a) shows the region where $\rho^+(\gamma, r) < 0$, while panel (b) illustrates the corresponding region for $\rho^-(\gamma, r)>0$.}
\label{fig2}
\end{figure}

This equation is solvable for $\omega^{\pm}_{t}$ in terms of $\beta$. The possible solutions for  $\omega^{+}_{t}$ and  $\omega^{-}_{t}$ are presented in Tables \ref{Tab1} and \ref{Tab2}, respectively. These values are obtained by numerically solving Eq.~(\ref{s15}) across allowed ranges of the EoS parameters, after which the corresponding $\beta$ and $m$ values are derived via consistency relations. The digits shown are rounded to the precision necessary to accurately reflect both qualitative and quantitative trends. Once the equation is solved for $\beta$, $m$ can be determined using Eq. \ref{26}.
A positive energy density is a necessary physical condition for the viability of wormhole solutions. Under this condition, it is evident that $H > 0$ holds for $\omega_r > -1$, and $H_1 > 0$ holds for $\omega_t > -1$. Applying this condition and the condition of asymptotic flatness reveals that the solutions for $\omega^{+}_t$ fail to meet this requirement; consequently, they are not physically relevant. The second category of solutions for $\omega^-_{t}$ is presented in Table \ref{Tab2}. Next, we examine the solutions corresponding to $\omega^{-}_{t}$, focusing specifically on values of $\beta$ that yield asymptotically flat solutions. Setting $\beta = -0.47520$ in (\ref{24d}) results in
\begin{equation}\label{s16}
 \rho_{1}^{\pm}=\frac{0.4984}{\gamma} \left(-0.5543\pm\sqrt{0.3073-\gamma\alpha\frac{7.0845}{r^{4.7657}}} \right),
\end{equation}
where
\begin{equation}\label{s17}
\Delta=0.3073-\gamma\alpha\frac{7.0845}{r^{4.7657}}
\end{equation}
A positive $\Delta$ is required to keep the energy density real.
As $\Delta$ is a general function of $r, \alpha,$ and $\gamma$, analyzing its behavior—and by extension, that of $\rho$—across the entire parameter space is highly complex. To simplify this investigation, we fix one parameter to examine the relationships between the others. We begin by setting $\gamma = 1$ so
 \begin{equation}\label{s16a}
\rho_{1}^{\pm}=0.4984 \left(-0.5543\pm\sqrt{0.3073-\alpha\frac{7.0845}{r^{4.7657}}} \right),
\end{equation}
where
 \begin{equation}\label{s17a}
\Delta=0.3073-\alpha\frac{7.0845}{r^{4.7657}}
\end{equation}
It is evident that $\Delta$ reaches its maximum value at $r=r_0=1$. Solving $\Delta(r=1)=0$ yields $\alpha = 0.0433$. Consequently, the condition $\Delta > 0$ is satisfied for
 \begin{equation}\label{s18}
  \alpha \leq 0.0433.
  \end{equation}
Next, we examine the ECs for this shape function. We begin by investigating the energy density across various ranges of $\alpha$ and $r$. In Fig. \ref{fig1}, we plot the two branches of the energy density, $\rho^+$ and $\rho^-$, using the same model parameters. Panel (a) shows the $\rho^+$ branch, while panel (b) shows the $\rho^-$ branch; the parameter values themselves remain unchanged between the two panels.
As illustrated in Fig. \ref{fig1}, $\rho^{+}(\alpha, r)$ takes negative values, indicating that it does not represent a viable wormhole solution; conversely, $\rho^{-}(\alpha, r)$ remains positive. By setting $\omega_r=0.6546$ and $\omega_t=-0.0647$, we demonstrate that all ECs are satisfied for $\beta =- 0.47520$ when considering the $\rho^{-}$ case.

 Next, we hold $\alpha$ constant at $\alpha=1$ and examine the wormhole solutions as a function of the variable $\gamma$. In this case,
\begin{equation}\label{s16ab}
 \rho_{1}^{\pm}=\frac{0.4984}{\gamma} \left(-0.5543\pm\sqrt{0.3073-\gamma\frac{7.0845}{r^{4.7657}}} \right),
\end{equation}
where
 \begin{equation}\label{s17ab}
\Delta=0.3073-\gamma\frac{7.0845}{r^{4.7657}}.
\end{equation}
Setting $\Delta(r=1) = 0$ results in $\gamma = 0.043$; thus, the condition $\Delta > 0$ holds for
\begin{equation}\label{s18a}
  \gamma \leq 0.0433.
  \end{equation}
An analysis of the energy density across varying ranges of $\alpha$ and $r$ (Fig. \ref{fig2}) reveals that $\rho^+(\gamma, r)$ assumes negative values, rendering it an unphysical wormhole solution. In contrast, $\rho^-(\gamma, r)$ remains positive throughout. By setting $\omega_r = 0$.6546 and $\omega_t = -0.0647$, we can show that all ECs are satisfied for $\beta = -0.47520$ within the $\rho^-$ configuration.
\begin{figure}
\subfloat[ $\rho^+<0$ ]{\includegraphics[width = 2.5in]{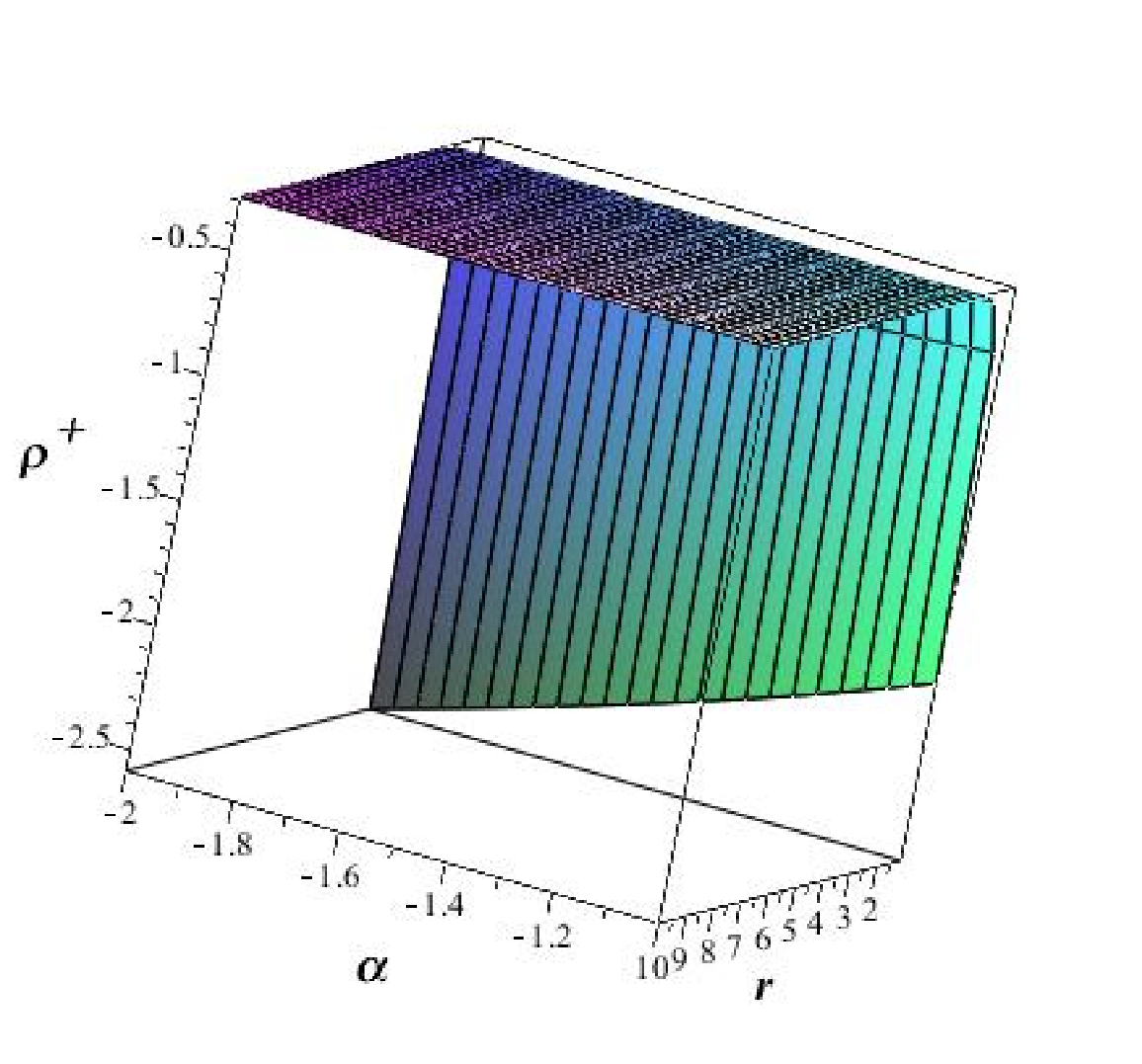}}\\
\subfloat[ $\rho^->0$]{\includegraphics[width = 2.5in]{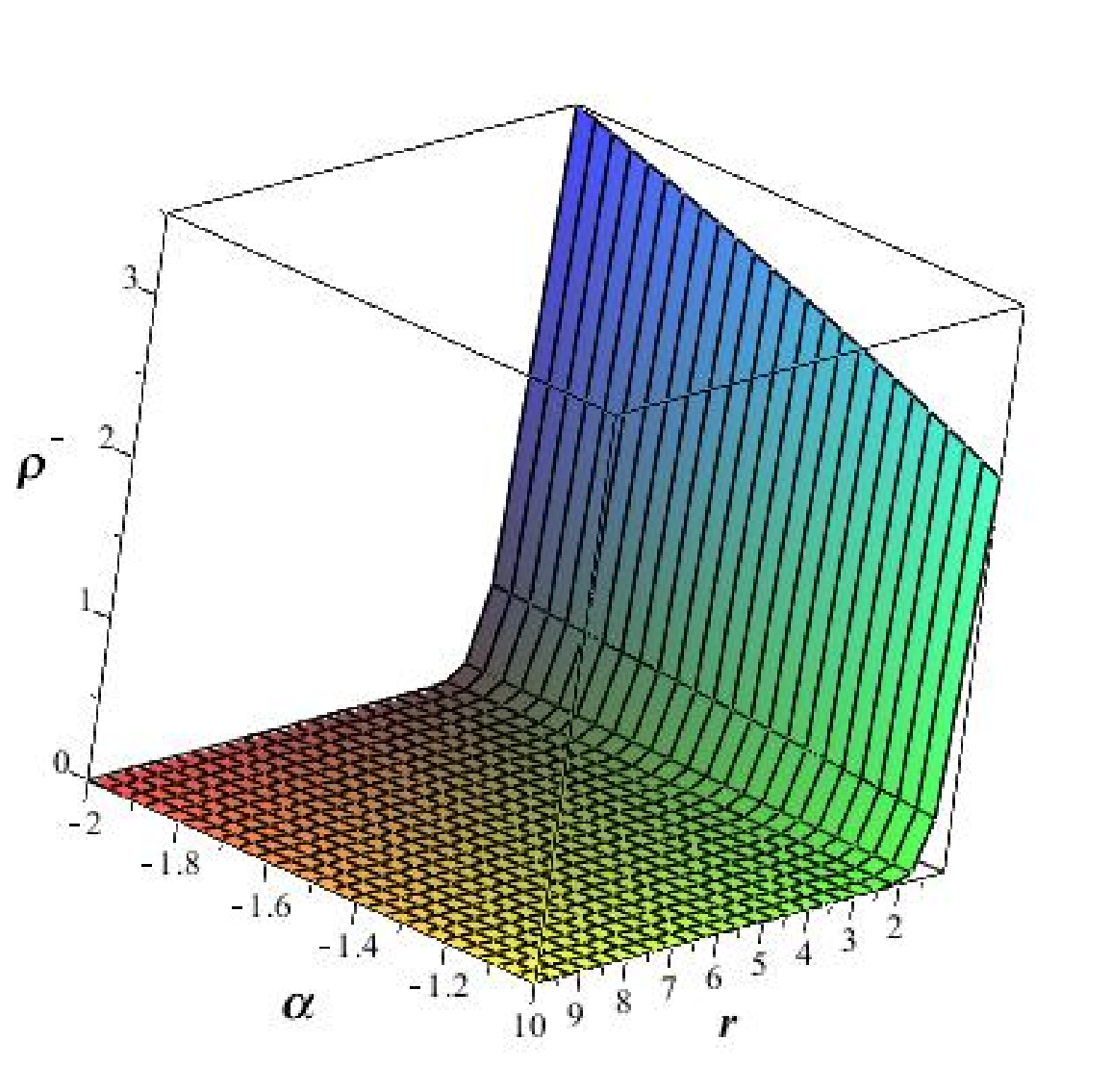}}
\caption{The relationship between $\rho^{\pm}(\alpha, r)$ and the variables $\alpha$ and $r$ for $\beta=-0.75309$ and $m=-7.686$. Panel (a) shows the region where $\rho^+(\alpha, r) < 0$, while panel (b) illustrates the corresponding region for $\rho^-(\alpha, r)>0$.}
\label{fig3}
\end{figure}

\begin{figure}
\subfloat[ $\rho^+<0$ ]{\includegraphics[width = 2.5in]{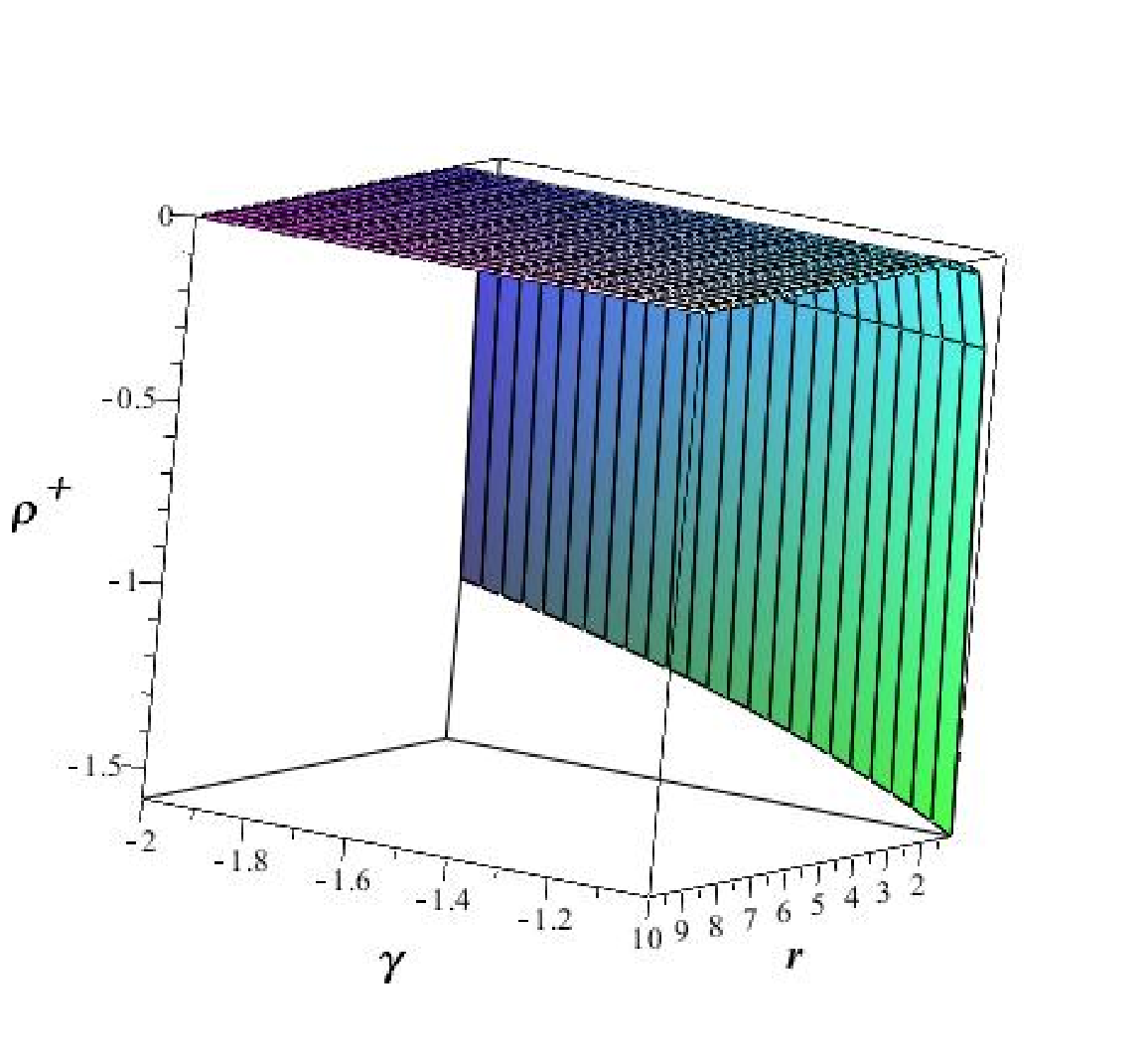}}\\
\subfloat[ $\rho^->0$]{\includegraphics[width = 2.5in]{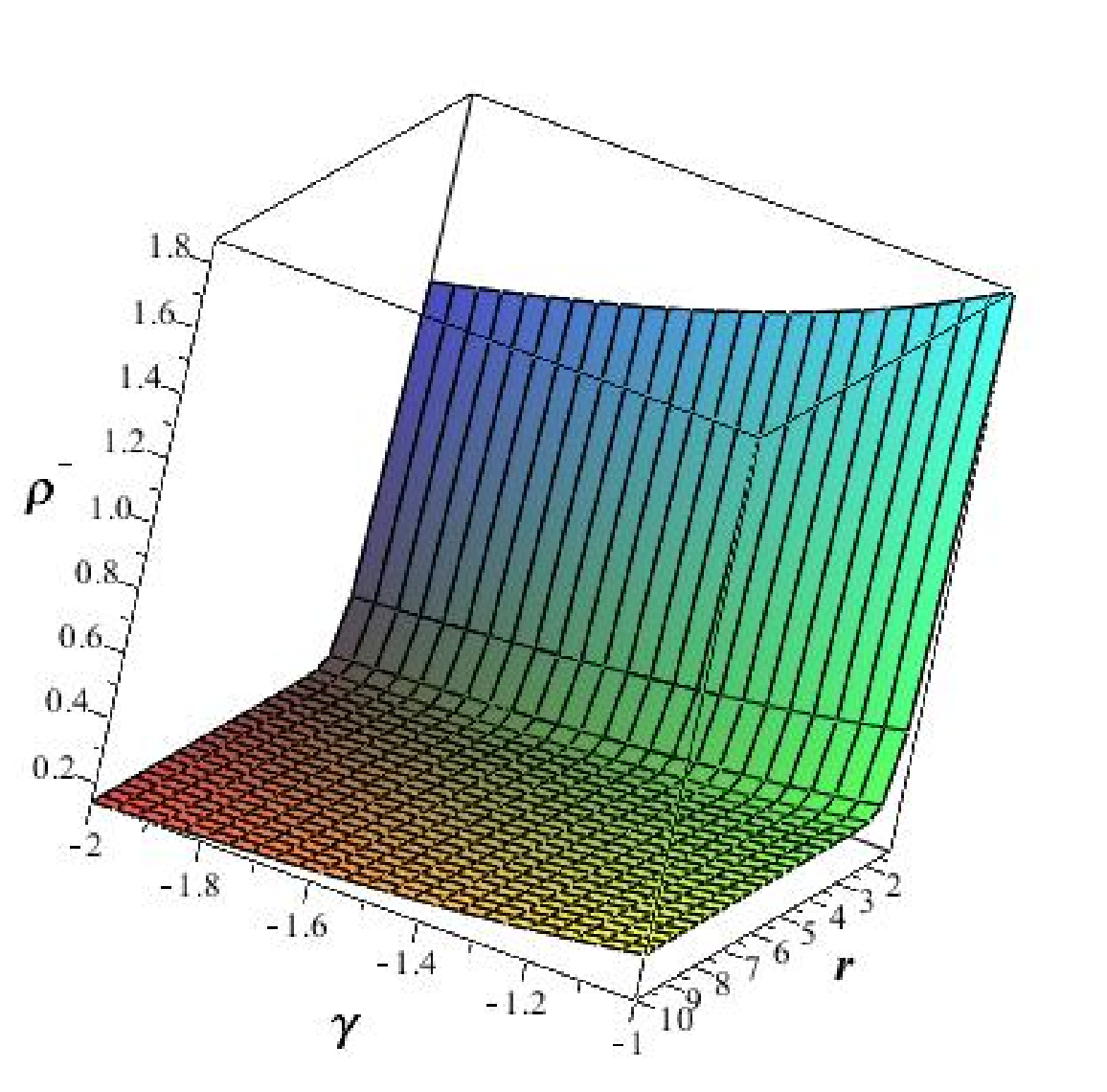}}
\caption{The relationship between $\rho^{\pm}(\gamma, r)$ and the variables $\gamma$ and $r$ for $\beta=-0.75309$ and $m=-7.686$. Panel (a) shows the region where $\rho^+(\gamma, r) < 0$, while panel (b) illustrates the corresponding region for $\rho^-(\gamma, r)>0$.}
\label{fig4}
\end{figure}

The same analyses can be applied for $\beta = 0.75309$. By setting $\gamma=1$, one reaches
\begin{equation}\label{s16ac}
\rho_{1}^{\pm}=0.5351 \left(-0.2669\pm\sqrt{0.0712-\alpha\frac{28.72}{r^{10.685}}} \right),
\end{equation}
where
 \begin{equation}\label{s17ac}
\Delta=0.0712-\alpha\frac{28.72}{r^{10.685}}
\end{equation}
Setting $\Delta(r=1) = 0$ yields $\alpha = 0.0024$, which implies that the condition $\Delta > 0$ is satisfied for
\begin{equation}\label{s18b}
  \alpha \leq 0.0024.
  \end{equation}
Figure \ref{fig3} illustrates the dependence of $\rho^{+}(\alpha, r)$ and $\rho^{-}(\alpha, r)$ on the parameters $\alpha$ and $r$. The results demonstrate that $\rho^{-}(\alpha, r)$ is physically acceptable.

 For $\alpha=1$, one can show that the condition $\Delta > 0$ is satisfied for
\begin{equation}\label{s18c}
  \gamma \leq 0.0024.
  \end{equation}
Figure \ref{fig4} depicts the influence of parameters $\gamma$ and $r$ on $\rho^{+}(\gamma, r)$ and $\rho^{-}(\gamma, r)$. The results confirm that $\rho^{-}(\gamma, r)$ satisfies physical acceptability criteria. The results in Table \ref{Tab2} demonstrate that $\beta = -0.7530$ satisfies all ECs for $\rho^{-}$ across certain ranges of $\alpha$ and $\gamma$.

\section{Concluding remarks}\label{sec4}

The concept of a wormhole has long fascinated both theoretical physicists and science fiction enthusiasts. However, the physical realization of a traversable wormhole faces a fundamental hurdle: the requirement for "exotic matter" that violates the classical ECs. 

The $f(Q, T)$ gravity represents a bold shift in how we model the universe: it moves away from the "curvature" of GR toward "non-metricity," while suggesting that matter and the very fabric of spacetime are inextricably linked in a way that standard physics does not account for. It remains one of the most promising "laboratories" for testing alternatives to Einstein's masterpiece.

The linear model (\ref{99a})  is the simplest extension of Symmetric Teleparallel Gravity. Because the dependence on $T$ is linear, the coupling is relatively "stiff." The field equations retain a structure very similar to GR, with $\alpha$ acting as a scaling factor for the gravitational constant.
Adding the $\gamma T^2$ term introduces non-linear coupling between matter and geometry, which significantly changes the dynamics. While wormhole geometries have been extensively studied within the framework of $f(Q, T)$ gravity using linear models, this study explores non-exotic, asymptotically flat wormhole solutions under more general conditions. Due to the increased complexity of the field equations compared to the linear case, we assume a linear EoS for the radial and transverse pressures. This approach yields three non-linear equations for the energy density ($\rho$).

Since each equation provides two potential solutions, we impose a consistency requirement to ensure they coincide. Our derivation yields two branches for the energy density ($\rho^\pm$). Through this analytical constraint, we derive a power-law shape function ($b(r)=r^m$). Applying the necessary mathematical constraints (reality and positivity of  $\rho$) yields specific relations between physical quantities such as $\omega_r$, $\omega_t$, $m$ and $\beta$. Our analysis indicates that these solutions are restricted to finite values of $\beta$, which in turn limits the possible values of $m$(see Tables (\ref{Tab1}) and (\ref{Tab2})). Further investigation reveals that many of these solutions are either not asymptotically flat or violate ECs. Consequently, non-exotic, asymptotically flat wormhole solutions ($m<1$) are highly constrained. Finlay, our analysis indicates that the $\rho^-$ branch is the physically acceptable solution.  This suggests that introducing non-linear terms into the $f(Q, T)$ function complicates the field equations and significantly restricts the domain of viable wormhole solutions. Our investigation reveals how the model parameters $\alpha$, $\beta$, and $\gamma$ influence both the shape functions and the existence of non-exotic solutions. Furthermore, our mathematical analysis establishes specific constraints on these parameters.

In $f(Q,T)$ gravity, the coupling between non-metricity and matter introduces additional degrees of freedom, modifying the gravitational field equations. These generate effective stresses capable of supporting wormhole throats without relying on fundamentally exotic matter. Consequently, $f(Q,T)$ wormholes provide critical insights into how geometry-matter coupling influences spacetime topology, ECs, and the viability of traversable solutions. The "exotic matter" requirement is arguably the greatest weakness of the classical wormhole model.
Within the specific $f(Q,T)=\alpha Q+\beta T+\gamma T^2$ model and for the restricted parameter regions identified in this work, the physical matter sector can satisfy the standard ECs while the modified geometric contributions participate in supporting the wormhole geometry. This does not constitute a theory-independent statement that modified gravity universally removes exoticity.


\end{document}